\documentclass[twocolumn,aps]{revtex4}
\input psfig.tex 
\newcommand{\mnras}{MNRAS}
\newcommand{\aap}{A\&Ap}
\newcommand{\physrep}{Phys.~Rep.}
\newcommand{\beq}{\begin{equation}}
\newcommand{\eeq}{\end{equation}}
\newcommand{\eea}{\end{eqnarray}}
\newcommand{\bea}{\begin{eqnarray}}
\newcommand{\HD}{H_{\rm DGP}}
\newcommand{\Hs}{H_{\rm std}}
\newcommand{\ml}{M_{\rm lim}}

\begin{document}

\title{Probing Modified Gravity by Combining Supernovae and Galaxy Cluster Surveys}
\author{Jia-Yu Tang, Jochen Weller and Alan Zablocki}

\affiliation{University College London, Gower Street, London, WC1E 6BT, U.K.}

\date{\today}

\begin{abstract}
Possible explanations of the observed accelerated expansion of
the Universe are the introduction of a dark energy component or the
modifications of gravity at large distances. A particular difference
between these scenarios is the dynamics of the growth of structures.
The redshift distribution of galaxy clusters will probe this growth of
structures with large precision. Here we will investigate how proposed
galaxy cluster surveys will allow one to distinguish the modified
gravity scenarios from dark energy models. We find that cluster counts
can distinguish the Dvali-Gabadadze-Porrati model from a dark energy
model, which has the same background evolution, as long as the
amplitude of the primordial power spectrum is constrained by a CMB
experiment like Planck.
In order to achieve this, only a couple
of hundred clusters in bins of width $\Delta z = 0.1$ are
required. This should be easily achievable with forthcoming
Sunyaev-Zel'dovich cluster counts, such as the South Pole Telescope in
conjunction with the Dark Energy Survey.
\end{abstract}
\maketitle
\section{Introduction}
During the course of the last ten years evidence for an accelerated
expansion of the Universe has been mounting. Initially driven by
Type Ia Supernovae (SNe) observations by the Supernovae Cosmology Project (SCP) and
the High-z Supernovae Search Team \cite{Perlmutter:97,Perlmutter98,Riess:98}, the
combination of large scale structure surveys like the 2 degree field
(2dF) and the Sloan Digital Sky Survey (SDSS) \cite{Percival01,Abazajian:03}, cosmic
microwave background (CMB) \cite{Spergel03,Spergel:06} and x-ray observations of clusters
of galaxies \cite{Allen:02,Rapetti:04} confirmed these findings. In recent
years even larger samples of Type Ia SNe \cite{Knop:03,Riess:04,Astier:06}
indicate strongly that the expansion of the Universe is speeding up. While
the observations are getting better and better, theoretical models of
accelerated expansions are still in its infancy. The most natural
candidate seems to be a cosmological constant, however this comes with
a drawback of fine tuning of the initial conditions to unnatural values. 

In general there are three possible approaches to obtain accelerated
expansion of the Universe. Firstly there could be an additional 
component in the energy-momentum tensor, which dominates the late time
dynamics of the Universe and leads to accelerated expansion. This component requires negative pressure and an example
of such a fluid could be a nearly homogeneous scalar field prevailing
in the Universe \cite{Wetterich:88, Peebles:88, Ratra:88, Ferreira:98, Zlatev:98}. The second possibility, is that Einstein's equations
of gravity require modification on large scales \cite{Dvali:00,Carroll:04,Capozziello:05,Carroll:05,Lue:06}. The third possible
explanation is that the Universe is inhomogeneous on large scale,
although it is hard to explain the boundary conditions for such a
scenario in a natural way \cite{Kolb:05}. In this article we concentrate on the first
two possibilities.\\

Numerous dark energy models have been suggested in the past, where
canonical scalar fields with various potentials are only a sub-class
\cite{Wetterich:88, Peebles:88, Ratra:88,Ferreira:98,Zlatev:98}. In
all these models the background evolution can be
described by the equation of state, i.e. the relation between the
pressure and density of the dark energy fluid. The standard dark
energy models also have a standard evolution of the growth of
structures. Theories of modified gravity are only emerging as possible
explanations for the observed acceleration. One of the most prominent
models is one, which is motivated by brane cosmology involving
extra dimensions and hence lead to an effectively modified Friedman
equation on the 3+1 dimensional brane \cite{Dvali:00,Lue:06}. Another possibility is to have inverse
curvature terms in the Lagrangian of gravity. The additional terms
become relevant at the late time evolution of the Universe and can
lead to accelerated expansion and hence explain the Supernovae
data \cite{Carroll:04, Capozziello:05, Mena:06}. When
modifying gravity, one has to be very careful so as not to violate high
precision tests of gravity in the local Universe, like the Solar
System, and also not to introduce unphysical features in the theory, like
negative energy eigenstates, namely ghosts, or super-luminal modes \cite{Luty:04,Nicolis:05,Koyama:05,Lue:06,Chiba:03,Soussa:04,Navarro:05,Navarro:06,DeFelice:06}.
For the rest of this article we will concentrate on the
Dvali-Gabadadze-Porrati (DGP) model as an example for a modified gravity
model. The main reason for this is that in this model the dynamics of
linear perturbations has been worked out \cite{Song:05,Koyama:06a,Koyama:06}. Although, some care is required on large
scales \cite{Koyama:06,Sawicki:06}, this is not important for the cluster count
test, which probes power spectra on sub-horizon scales.

\section{Dark Energy versus DGP cosmology}
For the analysis presented here we take a DGP model as our fiducial
model. The DGP model is an example of a brane world model, where our
four dimensional Universe resides in a five dimensional space on a
brane \cite{Dvali:00}. The novelty about this model, compared to other
brane world models, is that the flat extra dimension is allowed to be
large\cite{Lue:06}. The changes of 
the cosmological evolution on the four dimensional brane, compared to
standard gravity, are caused by 'leakage' of gravitational degrees of
freedom off the brane. Cosmological solutions allow empty Universes
with accelerated expansion \cite{Deffayet:01,Deffayet:02}. The
resulting modified Friedman equation for the simplest of these models
is given by \cite{Lue:06}
\beq
\HD^2 - \frac{\HD}{r_c}=\frac{8\pi G}{3}\rho\, ,
\label{eqn:dgp}
\eeq
where $\HD$ is the Hubble parameter in the DGP model, $G$ the four
dimensional gravitational constant and $\rho$ the energy density of
the constituents of the Universe. For the analysis in this
article we focus on the late time evolution and set overall density $\rho$
equal to the matter density $\rho_{\rm
  m}$. Gravity on the brane is affected by 5-dimensional effects above the
crossover scale $r_c$. Self-acceleration of the Universe takes place
today if $r_c \sim H_0^{-1}$. For a
flat Universe, the parameter $r_c$ is related to the matter contents
$\Omega_{\rm m}$, by \cite{Lue:06}
\beq
\frac{1}{H_0 r_c} = 1-\Omega_{\rm m}\, ,
\label{eqn:dgpc}
\eeq
which ensures that one obtains $H_0$ as the expansion rate of the
Universe today.

In order to compare the evolution of Universe in the DGP world with
the one of a dark energy model, we note that the Hubble parameter in
standard gravity at late times is given by
\beq
\Hs^2 = \frac{8\pi G}{3}\left\{\rho_{{\rm m},0}a^{-3}+\rho_{{\rm
    de},0}\exp\left[-3\int_1^a\frac{1+w(a')}{a'}\,da'\right]\right\}\;,
\label{eqn:std}
\eeq
with $w(a)$ the equation of state factor of the dark energy
component. By identifying the Hubble parameters in
Eqn. (\ref{eqn:dgp}) and (\ref{eqn:std}) we obtain $w(a)$ 
which mimics the background evolution of the DGP model
\beq
w(a) = -1 + \frac{\Omega_{\rm m}a^{-3}}{\left[(r_c
    H_0)^{-1}+2\eta\right]\eta}\; ,
\label{eqn:wa}
\eeq
with $\eta = \sqrt{\Omega_{\rm m}a^{-3}+1/(2r_c H_0)^2}$. This tends
to $w(a=1) = -1/(1+\Omega_m)$ today and to $w(a=0) = -1/2$ at early
times, if we include the condition in Eqn. (\ref{eqn:dgpc}).
\begin{figure}[!h]
\setlength{\unitlength}{1cm}
\centerline{
\hbox{\psfig{file=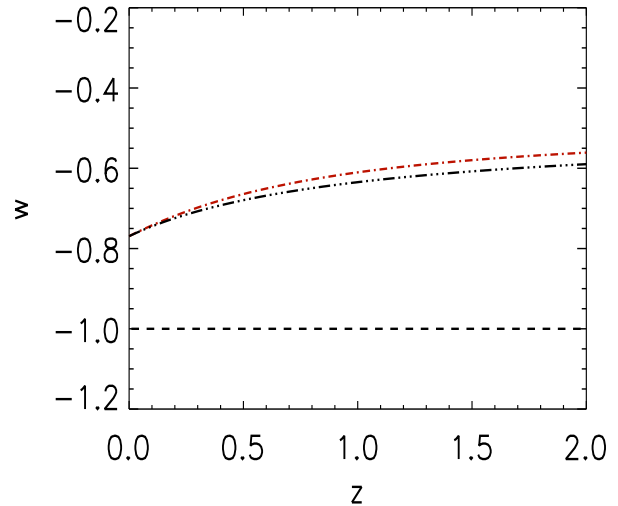,height=8cm,width=9cm}}} 
\caption{The mimicking $w(z)$ for the DGP model (dot-dashed line). The
  triple dot-dashed line is the 'best fit' parameterization with $w(z) =
  w_0+w_az/(1+z)$. We find $w_0 = -0.77$ and $w_a=0.27$ for the
  fiducial DGP model.}
\label{fig:wa}
\end{figure}
In Fig.~\ref{fig:wa} we show the mimicking equation of state factor
(dot-dashed line), if we set $\Omega_{\rm m} = 0.3$. Given that a
$\Lambda$CDM model ($w=-1$, dashed line) fits current observations
very well, this model might be already under some strain from
observations. The behaviour of the mimic model at late and early
times also allows to obtain the coefficients of a popular
parameterization of the equation of state factor, which is given by
$w(a)=w_0+w_a(1-a)$ \cite{Chevallier:01,Linder:03}. For the case we
discuss here the coefficients are: $w_0 = -0.77$ and $w_a = 0.27$. The
equation of state factor for this parameterization is shown by the triple
dot-dashed line in Fig.~\ref{fig:wa}. 

We will now address the question how a purely geometrical probe, like
the Supernovae magnitude - redshift relation, can constrain the DGP
model. For this we calculate the model predictions for the
magnitudes, which are given by $m \propto \log d_{L}$ and the luminosity
distance $d_L = (1+z)\int_0^z dz'/H(z')$. 
\begin{figure}[!h]
\setlength{\unitlength}{1cm}
\centerline{
\hbox{\psfig{file=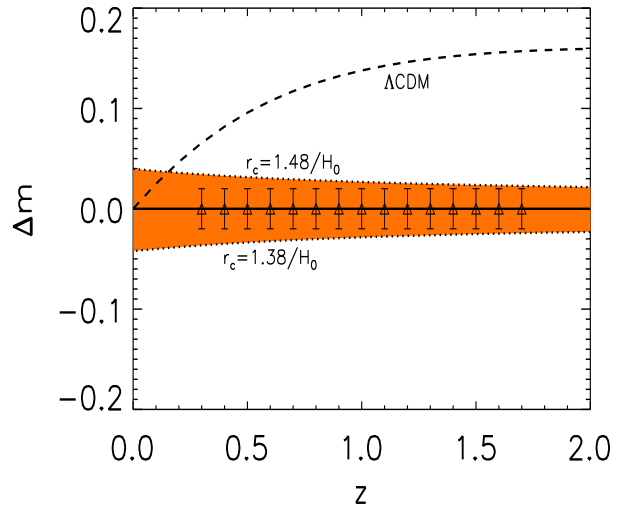,height=8cm,width=9cm}}} 
\caption{Magnitude-redshift relation with respect to the fiducial DGP
  model with $r_c = 1.43/H_0$ (solid). The shaded region is
  for models where $r_c$ varies from $1.38/H_0$ to $1.48/H_0$. The
  errorbars are the statistical errors from a SNAP-like
  experiment. The dashed line is for $\Lambda$CDM Universe.} 
\label{fig:sne}
\end{figure}
In Fig.~\ref{fig:sne} we plot the magnitude difference to the fiducial
DGP model with $H_0 = 72 {\rm km}/{\rm s}/{\rm Mpc}$ and $\Omega_{\rm m} = 0.3$. From
this we obtain with Eqn.~\ref{eqn:dgpc} a fiducial value for $r_c H_0 \approx 1.43$. The shaded
region shows the magnitude difference if we vary $r_c H_0$ between
$1.38$ and $1.48$. Note that for these values
the Hubble parameter today is still well within the errorbars of its
measured value of $H_0 = (72\pm 8) {\rm km}/{\rm s}/{\rm
  Mpc}$ \cite{Freedman01}. We also include mock data points from a
SuperNovae Acceleration Probe (SNAP)-like
survey \cite{SNAP:05a}. We assume a distribution of 2000 Type Ia SNe between
$z=0.3$ and $z=1.7$ with a magnitude error of $\sigma_m = 0.15$ for an
individual SNe \cite{Linder:03b}. We plot the statistical errorbars in redshift bins of
$\Delta z =0.1$. A SNAP like experiment will not be able to
distinguish DGP models with $1.38<r_c H_0<1.48$. Realistically the
situation is worse, since such large bins probably hit the systematic
error limit of the survey. Nevertheless, it is evident from the plot,
that SNAP could distinguish the DGP models from a $\Lambda$CDM model
(dashed line). Of course the dark energy model with an equation of
state given by Eqn.~(\ref{eqn:wa}), would be spot on top of the
fiducial solid line in the plot. {\em A pure geometric probe, like the
  SNe magnitude-redshift relation, can
not distinguish a modified gravity model from an arbitrary dark energy
model.} However, a SNAP-like survey is likely to probe the geometry and hence the
background evolution very well.

So far we have concentrated on SNe as a cosmological probe and
established that SNe alone cannot distinguish dark energy from the
DGP model. However, recent studies have shown that weak lensing,
baryon oscillations or observations of the integrated Sachs-Wolfe
effect have potential to distinguish dark energy from modified gravity \cite{Song:05,Song:06a,Song:06,Yamamoto:06,Song:06b}. The
ability of these probes to distinguish the DGP model from dark energy
is driven by the different dynamics of the growth of structure. The
linear density perturbations in the DGP model are governed by \cite{Song:05,Koyama:06a,Lue:06,Yamamoto:06}
\beq
\delta
''+\delta'\left[\frac{3}{a}+\frac{H'}{H}\right]=\frac{3}{2}\frac{H_0^2}{H^2}a^{-5}\left(1+\frac{1}{3\beta}\right)\Omega_{{\rm
    m}}\delta\; ,
\label{eqn:pert}
\eeq
with $\beta = 1-\frac{2}{3}Hr_ca\left[3/a+H'/H\right]$, and primes
denoting derivatives with respect to the scale factor $a$. Note, that if we set
the term $1/(3\beta)=0$, we obtain the standard gravity result.
For the evolution of linear perturbations this is the key difference
between the DGP model and the mimic dark energy model.

\section{Galaxy Cluster Counts}
Counts of galaxy clusters are very sensitive to the evolution of the
linear growth factor. Clusters are thought to be formed in the peaks of
the linear density field, if the density is above a threshold density
of $\delta_c = 1.686$, for a flat matter dominated Universe. This is the
basis of the Press-Schechter approach for the distribution of clusters
in mass and redshift \cite{Press:74a}. The modern approach is to measure this
function with N-body simulations, which result in \cite{Jenkins:01a}
\bea
\frac{dn}{dM}\left(z,M\right) &=& - 0.316
\frac{\rho_{{\rm m},0}}{M}\frac{d\sigma_M}{dM}\frac{1}{\sigma_M} \nonumber \\
&& \times\exp\left\{-|0.67-\log\left[\delta(z)\sigma_M\right]|^{3.82}\right\}\, ,
\label{eqn:mass}
\eea
where $n$ is the number density of clusters of mass $M$, $\sigma_M$ is
the density fluctuation of a shell with mass M in terms of the mean
background mass density and $\delta(z)$ is the growth factor as a
function of redshift. It is clear from Eqn.~\ref{eqn:mass} that the
mass function is strongly dependent on the growth factor
$\delta$. Although, the mass function might be slightly different
 in modified gravity scenarios \cite{Maccio:04}, the
dominant contribution is the different dynamics of the growth of structures.
We will investigate if counts of galaxy clusters can distinguish the DGP
model from the mimic dark energy model. This is similar to the
earlier works, which concentrate on weak lensing, baryon oscillations and
integrated Sachs-Wolfe effect observations \cite{Song:05,Song:06a,Song:06,Yamamoto:06,Song:06b}. 

The number of clusters per redshift bin is given by \cite{Battye:03}
\beq
	\frac{dN}{dz} =
{\Delta\Omega}\frac{dV}{dzd\Omega}(z)\int\limits_{M_{\rm 
lim}(z)}^\infty \frac{dn}{dM}\,dM\, ,
\eeq
where $dV/(dzd\Omega)= [r(z)]^2/H$ is the comoving volume in a flat
universe, with $r(z)=\int_0^z H^{-1}(z^{\prime})dz^{\prime}$ the
coordinate distance and
$\Delta\Omega$ is the angular sky coverage of the survey. $\ml(z)$ is
the limiting mass, which will depend in general on the parameters of
the survey and cosmology.

The near future will see a few surveys, which will
perform a blind search for galaxy clusters and identify them via the
so called Sunyaev-Zel'dovich decrement in the CMB radiation \cite{Sunyaev:72}. The interesting feature of this
effect, is that it is independent of redshift and hence could
potentially identify clusters at large redshift. Upcoming surveys are
the South Pole Telescope (SPT) \cite{Ruhl:04}, Atacama Cosmology Telescope(ACT) \cite{Fowler:04} and
Atacama Pathfinder EXperiment (APEX) \cite{Gusten:06}. In order to perform number counts of
galaxy clusters in redshift bins we also require the redshift. The
proposed Dark Energy Survey (DES) \cite{DES} is designed to obtain the
redshifts of
the clusters identified by SPT. SPT will observe 4.000 ${\rm deg}^2$
of sky with a 1 arcminute beam. In order not to be too SPT specific in the
analysis presented here, we do not calculate the limiting mass of the
survey from the instrument parameters for SPT. We assume an {\em
  effective} constant mass limit of $\ml = 1.7\times 10^{14} h^{-1}
M_\odot$, which corresponds roughly to the mass limit of the SPT
survey \cite{Battye:03}. However the results we find will apply to any
kind of galaxy cluster survey. Apart from the cosmological parameters
mentioned so far, we also require for the analytical prediction of
the cluster counts the normalization and slope of the primordial
matter power spectrum. We choose a slope of $n=1$ and $\sigma_8 =
0.75$ \cite{Spergel:06}. For this survey and fiducial model we find
$\approx 7200$ clusters of galaxies. 
\begin{figure}[!h]
\setlength{\unitlength}{1cm}
\centerline{
\hbox{\psfig{file=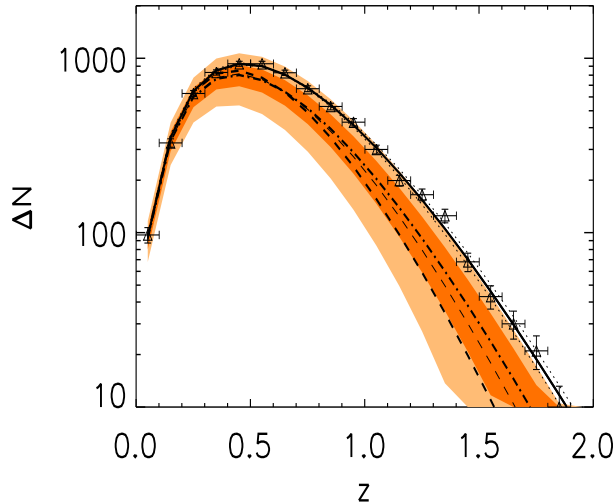,height=8cm,width=9cm}}} 
\caption{Number of clusters of galaxies above a threshold of $M_{\rm
    lim} = 1.7\times 10^{14}h^{-1}M_\odot$, for the concordance $\Lambda$CDM
  Universe (dashed) and the fiducial DGP model (solid). The thin
    dashed line is for a model with $w=-0.8$. The dotted lines are for
    the $r_c=1.38/H_0$ and $r_c=1.48/H_0$ models respectively. The
    dot-dashed line is for a dark energy model, which mimics the
    background evolution of the DGP scenario. The errorbars assume
  Poisson errors for each bin with $\Delta z = 0.1$ for a SPT-like
    experiment. The shaded regions are illustrating the uncertainties
    on the mimic model from propagating errors on $w_0$, $w_a$,
    $\Omega_{\rm m}$ and $\sigma_8$ from SNe and CMB observations. The
    light shaded region is obtained by assuming $\delta\sigma_8 =
    0.03$ and the dark shaded on for $\delta\sigma_8 = 0.01$.} 
\label{fig:dNdz}
\end{figure}
In Fig.~\ref{fig:dNdz} we show the results for counts of clusters in
redshift bins of $\Delta z = 0.1$. The solid line is the fiducial DGP
model, with randomly generated mock data points including Poisson
errors. The faint dotted lines on either side of the fiducial model
correspond to $1/(r_c H_0) = 1.38$ and $1/(r_c H_0) = 1.48$. Note that
these models are hardly distinguishable from the fiducial model in the
number counts. The dot-dashed line is for the mimic dark energy model
with equation of state given by Eqn.~\ref{eqn:wa}. The difference is,
particularly in the peak of the distribution, well above the $2\sigma$
threshold. This is driven by the difference in the growth
factor of the DGP model, given by Eqn.~\ref{eqn:pert} and the mimic
dark energy model, which is obtained for $\beta \to \infty$. So prospects are very good that galaxy cluster counts can
distinguish the modified gravity models from dark energy in
conjunction with a geometrical probe like SNAP. However, we also have
to address the problem of parameter degeneracies. It is very likely
that the mimic model with a different value of $\sigma_8$ is
degenerate with the DGP model. In order to study this effect we
perform a simplistic error propagation according to
$\delta\Delta N(z_i)^2 = \sum_{\{j,k\}}\frac{\partial\Delta
  N(z_i)}{\partial \theta_j}\frac{\partial\Delta N(z_i)}{\partial
  \theta_k}\delta\theta_j\delta\theta_k$,
with $\theta_j = \{w_0,w_a,\Omega_{\rm m},\sigma_8$\}. We adopt here
the parameterization $w(a) = w_0+w_a(1-a)$ of the mimic model with $w_0
= -0.77$ and $w_a = 0.27$. From a SNAP-like survey we expect
$\delta w_0 = 0.05$ and $\delta w_a = 0.2$ with $\delta \Omega_{\rm m}
= 0.03$ \cite{Linder:03b}. Current constraints on $\sigma_8$ from
combining WMAP3 and 2dF or SDSS are $\delta \sigma_8 = 0.03$
\cite{Spergel:06}. This case is illustrated as the light shaded region
in Fig.~\ref{fig:dNdz}. For Planck the constraint on the amplitude of the
primordial power spectrum is likely to improve by a factor 5
compared to WMAP \cite{Planck}. We hence adopt conservatively a prior
on $\sigma_8$ from Planck, possibly in addition to a large scale
structure measurement like SDSS, of $\delta\sigma_8 = 0.01$. This is
the dark shaded region in Fig.~\ref{fig:dNdz}. We see from the Figure
that even if we marginalize over parameter degeneracies we can still
distinguish the mimic model from the DGP model significantly above the
$1\sigma$ level.
One problem we have not addressed in this analysis, is the uncertainty
in the mass limit of the survey \cite{Battye:03,Lima:05}. This will
effectively increase the errorbars in Fig.~\ref{fig:dNdz}. However,
self-calibration methods \cite{Majumdar:04,Lima:05} and optical
calibration of the cluster masses  \cite{Majumdar:04,DES,Sealfon:06}
are very promising and could constrain the error in the mass limit. In
order to obtain some insight in how different the mimic dark energy
model is from the fiducial DGP model, we also plot the number count
prediction for the $\Lambda$CDM model (thick dashed line) versus a
model with $w=-0.8$ (thin dashed line). With self-calibration it
should be
possible to distinguish these two models \cite{Majumdar:04,DES},
particularly if priors from the Planck CMB mission on the other
cosmological parameters are available \cite{DES,Planck}. Given that the
difference between these two models is smaller than the one between
the DGP and mimic dark energy model, cluster counts should be able to
distinguish modified gravity from dark energy. 
\section{Conclusion}
We have shown that the background evolution of the DGP model can be
entirely mimicked by a dark energy model with a suitably chosen
equation of state, given in Eqn.~\ref{eqn:wa}. Due to this fact Type
Ia SNe alone can not distinguish modified gravity models from dark
energy. However, due to the different
time evolution of the linear perturbations, galaxy cluster counts can
distinguish the DGP model from a dark energy model with
the same background evolution. The difference between these two models
is more then at the $2\sigma$ level, even for relatively low number
of clusters $\approx 7000$. This result is only slightly weakened if
we include parameter degeneracies in the mimic model.
We have presented a very simplified analysis in the article, but the
striking difference between the mimic dark energy and the DGP model,
raises hopes that this result also stands up in a more careful
analysis, which for example includes uncertainties in the gas physics
of clusters in order to establish the mass limit \cite{Younger:06}. 

A further interesting question would be to see whether our findings also hold up
for other modified gravity scenarios, like the inverse curvature model
 \cite{Carroll:04}. In order to answer this question a perturbative
analysis of these models is required, which is not an easy
task. It would be very surprising if the standard perturbation
result holds for these models. In general we would expect that
modified gravity models lead to a modification of the growth of structure. Clusters of galaxies in combination with a geometrical probe,
like SNe distances, could provide a uniquely sensitive window for the
distinction between general modifications of gravity and dark energy.

{\bf Acknowledgement} We thank Filipe Abdalla and Richard Battye for
useful discussions.

\end{document}